\documentclass{ejasa}
\usepackage{threeparttable}
\usepackage{float}
\usepackage[english]{babel}

\newenvironment{sbmatrix}[1]
{\def\mysubscript{#1}\mathop\bgroup\begin{bmatrix}}
	{\end{bmatrix}\egroup_{\textstyle\mathstrut\mysubscript}}

\title{A mixed-frequency approach\\ for exchange rates predictions}

\newcommand{\authorlastnames}{Mattera, R. et al.}

\author[b]{Raffaele Mattera}

\author[a]{Michelangelo Misuraca\thanks{Corresponding author: michelangelo.misuraca@unical.it}}
\author[b]{Germana Scepi}
\author[b]{Maria Spano}

\affil[a]{University of Calabria - Department of Business Administration and Law\\
	address}
\affil[b]{University of Naples "Federico II" - Department of Economics and Statistics\\
     address}

\date{{\small \today }}

\newcommand{\ejasavolume}{00}
\newcommand{\ejasanumber}{00}
\newcommand{\ejasayear}{Xxxx 2021}
\newcommand{\ejasafirstpage}{1}
\newcommand{\ejasalastpage}{18}
\newcommand{\ejasadoi}{DOI--xxxx}

\begin{document}






 \maketitle 
\setcounter{page}{\ejasafirstpage}


\begin{abstract}
Selecting an appropriate statistical model to forecast exchange rates is still today a relevant issue for policymakers and central bankers. The so-called \emph{Meese and Rogoff puzzle} assesses that exchange rate fluctuations are unpredictable. In the literature, a lot of studies tried to solve the puzzle finding alternative predictors and statistical models based on temporal aggregation. In this paper, we propose an approach based on mixed frequency models to overcome the lack of information caused by temporal aggregation. We show the effectiveness of our approach in comparison with other proposed methods by performing CAD/USD exchange rate predictions.
\paragraph{keywords:} MIDAS, linear regression, frequency alignment, forecasting
\end{abstract}

\section{Introduction}

According to a popular quote attributed to Alan Greenspan, the former U.S. Federal Reserve Chairman: ``\emph{implicit in any monetary policy action or inaction is an expectation of how the future will unfold, that is, a forecast'}' \citep{carlstrom1999}. Exchange rate forecasting is an essential issue for policymakers and central bankers because these predictions are used to project in the future the potential consequences of given monetary policies. Central bank policies in the U.S. and EU are described by interest rate rules, where interest rates respond to forecasts of inflation and economic activities rather than just outcomes \citep{wieland2013}. Equally important, exchange rate predictions result extremely decisive for heavy importer/exporter countries' central banks.

Several choices have to be made to forecast the exchange rates. First of all, a set of significant predictors has to be defined. Economic theory \citep[e.g.][]{fisher1896, frenkel1976, choi2003} provides a powerful guide. Among \emph{classical} theoretical frameworks, it can be mentioned the interest rate differential (\emph{uncovered interest rate parity} theory), the price levels differential (\emph{purchaising power parity} theory) and the money supply (\emph{monetary} theory). A subsequent but equally relevant aspect in forecasting is the time-frequency. Some studies on exchange rates focused on monthly predictions \citep[e.g.][]{molodtsova2009}, whereas other studies aimed at forecasting exchange rates with quarterly predictions \citep[e.g.][]{cheung2005, cheung2019}. Time horizon is an important choice since there is an interest in obtaining either short-run and long-run forecasts. A third aspect concerns the selection of a statistical model. Models could either be based on single or multiple relations, with linear or nonlinear specifications, allowing or not allowing co-integration by considering an error correction term \citep[see][]{rossi2013}.

The so-called \emph{Meese and Rogoff puzzle} \citep{meese1983, meese1988} assesses that, differently from what is claimed by the economic theory, exchange rate fluctuations are challenging to predict in practice. As a main result, the simple random walk model provides more accurate forecasts than the most competing models based on classical predictors. Previous studies tried to solve the puzzle by finding new competing predictors and statistical models to forecast exchange rates better than the random walk model. Concerning the choice of the variables, \cite{meese1988} tested classical predictors' forecasting accuracy against the random walk hypothesis. A similar approach was proposed by \cite{mark1995,chinn1995,cheung2005,cheung2019}. \cite{molodtsova2009} showed that the Taylor rule-based variables are, to some extent, able to forecast the exchange rates. Similarly, \cite{ferraro2015} showed that oil price fluctuations play an important role at this aim. As regards the choice of statistical models, instead, \citeauthor{meese1988} used the classical linear regression to obtain predictions, whereas \citeauthor{mark1995} proposed long-run relationships among predictors and exchange rates with \emph{error correction models} (ECM). Nevertheless, this approach showed several drawbacks in forecasting exchange rates. The puzzle is still not properly solved, and some questions remain open. For example, it is not clear why with monthly data some authors as \citeauthor{molodtsova2009} and \citeauthor{ferraro2015} had evidence in favour of predictability, while other authors \citeauthor{cheung2005} obtained a favourable result with quarterly data. Among the others, \cite{meese1988} tried to explain the puzzle trough sampling errors or model misspecification.

Exchange rate data are daily available, and all the related forecasting studies used a data aggregation step. This operation induces the so-called \emph{temporal aggregation bias} \citep{marcellino1999}, consisting of a considerable loss of information once data aggregation is used. In this paper, we aim at exploring the role of temporal aggregation bias in the Meese and Rogoff puzzle. The challenge is how to handle the mixture of sampling frequencies in exchange rates' predictions. For this purpose, we implemented a strategy based on the so-called \emph{Mixed Data Sampling regression} \citep[MIDAS:][]{foroni2015}, which allows analysing data with different time frequency.

The paper is structured as follows. In Section 2, we briefly reviewed the predictors used to forecast exchange rates by previous studies. Then, in Section 3, we described the implemented statistical methodology. In Section 4, we provided empirical evidence of the forecasting ability of mixed frequency models, presenting a case study related with CAD/USD exchange rate prediction. We closed the paper with some remarks.

\section{Classical predictors for exchange rates}

In exchange rates' forecasting, the class of theoretical models that have been tested over time against the random walk hypothesis is vast. The selected benchmark, the random walk without drift, could be written as:

\begin{equation}\label{eq:0}
\Delta s_{t} = \Delta s_{t-1} + \epsilon_t
\end{equation} 
\\
where $\Delta s_t$ is the exchange rate differential and $\epsilon_t$ an error term. In this section, we briefly examine the most relevant models used into the reference literature.

\subsection{Uncovered Interest Rate Parity}

According to the \emph{uncovered interest rate parity} (UIRP) theory \citep{fisher1896}, the interest rate differentials between two countries should explain fluctuations in the exchange rates. However, many previous studies showed that more accurate forecasts can be obtained by using the random walk. Several authors found good results using monthly frequency data \citep{clark2006,molodtsova2009}. The UIRP model could be specified by estimating the following equation:

\begin{equation}\label{eq:1}
\Delta s_{t} = \alpha + \beta (i_t - i^*_t) + \epsilon_t
\end{equation}
\\
where $\alpha$ is the intercept term of the model, $i_t$ is the domestic short-term interest rate, $i^*_t$ the foreign short-term interest rate, $\beta$ is the relation between the interest rate differentials and the exchange rate fluctuations and $\epsilon_t$ is the error term. Some studies restricted $\alpha=0$. A positive value of $\beta$ produces a forecast of the interest rate depreciation.

\subsection{Purchasing Power Parity}

Another classical predictor could be find in the \emph{purchasing power parity} theory (PPP) considering the price level differentials of two countries. In particular, to test the validity of the PPP theory for exchange rate forecasting, the following equation is estimated:

\begin{equation}\label{eq:2}
\Delta s_{t} = \alpha + \beta (p_t - p^*_t) + \epsilon_t
\end{equation}
\\
where $p_t$ is the domestic price level, $p^*_t$ the foreign price level, $\alpha$ and $\beta$ the parameters to be estimated and $\epsilon_t$ the error term. As above, $\alpha$ could either be restricted to zero or not. Previous studies showed that the out-of-sample performances are not good for the PPP model. In particular, \cite{cheung2005} found that predictors based on PPP produce more accurate forecasts than random walk within a long timescale but their performance are never significantly better. \cite{molodtsova2009} showed instead that the PPP model is significantly worse than a random walk in shorter time horizons. Similar results can be found in \cite{cheung2019}.

\subsection{Monetary models}

The monetary models \cite[e.g.][]{frenkel1976} assess that the exchange rates are determined by the movements in countries' relative money supply, outputs, interest rates and prices. Assuming that UIRP and PPP hold, the following equation is estimated:

\begin{equation}\label{eq:3}
\Delta s_{t} = \alpha + \beta_1 (i_t - i^*_t) + \beta_2 (y_t - y^*_t) + \beta_3 (m_t - m^*_t) + \epsilon_t
\end{equation}
\\
where $y_t$ and $m_t$ are the output and the money supply, respectively. The $\beta_3$ coefficient on money differentials is usually restricted to 1, whereas $\beta_2$ is assumed to be negative since $(y_t - y^*_t) < 0$ implies a domestic currency depreciation with an increasing (\emph{ceteris paribus}) foreign country output. Equation \ref{eq:3} has been defined \emph{flexible price version of the monetary model} by \cite{meese1988}. 

Another monetary model is the so called \emph{sticky price}, where it is supposed that the PPP holds only in the long run. The main difference with respect to the specification in Eq. \ref{eq:3} is that the functional relation is enriched by the price levels' differentials:

\begin{equation}\label{eq:4}
\Delta s_{t} = \alpha + \beta_1 (i_t - i^*_t) + \beta_2 (y_t - y^*_t) + \beta_3 (m_t - m^*_t) + \beta_4 (p_t - p^*_t) + \epsilon_t
\end{equation} 
\\
Even if \cite{mark1995} found strong and statistically significant evidence in favour of these predictors in a very long time horizon (three to four years), these results have been later argued by several scholars \cite[e.g.][]{chinn1995, cheung2005, molodtsova2009, cheung2019}. \cite{meese1983} demonstrated that the random walk is better than any monetary models in forecasting exchange rates. These findings have been confirmed by \citeauthor{chinn1995} for a short timescale, by \citeauthor{cheung2005} for very long horizon time (five years) and by \citeauthor{molodtsova2009}, which found good evidence just for few countries.


\subsection{Taylor rule fundamentals}

Some authors proposed to use predictors based on the \emph{Taylor rule} of monetary policy \citep{taylor1993} to forecaste the exchange rates. Taylor theorised that monetary authorities set the real interest rate as a function of how inflation differs from a given target. According to this claim, the central banks' response function can be expressed as:

\begin{equation}\label{eq:5}
\hat{i}_t = \pi_t + \phi (\pi_t - \bar{\pi}) + \gamma y^{gap}_t + \bar{r}
\end{equation}
\\
where $\hat{i}_t$ is the target short-term interest rate, $\pi_t$ is the inflation rate at current time, $(\pi_t - \bar{\pi})$ is the deviation of the current inflation rate from its target level $\bar{\pi}$, $y^{gap}_t$ is the output gap and $\bar{r}$ is the equilibrium level of real interest rate. The parameters $\phi$ and $\gamma$ define how the inflation rate and the output gap affect the target interest rate. Following \citeauthor{molodtsova2009}, we can combine $\pi_t$ and $\bar{r}$ into a constant term such that:

\begin{equation}\label{eq:6}
\hat{i}_t = \mu_t + \phi \pi_t + \gamma y^{gap}_t
\end{equation}
\\
where $\mu_t = \bar{r} - \phi \bar{\pi}$. The same relation hold for a foreign central bank:

\begin{equation}\label{eq:7}
\hat{i}^*_t = \mu^*_t + \phi \pi^*_t + \gamma y^{*gap}_t
\end{equation}
\\
Assuming that the interest rate $i_t$ immediately reaches the target $\hat{i}_t$, and that both central banks set the interest rates according to a Taylor rule, if the UIRP holds we get:

\begin{equation}\label{eq:8}
\Delta s_{t} = \alpha + \beta_1 (\pi_t - \pi^*_t ) + \beta_2 (y^{gap}_t - y^{*gap}_t) + \epsilon_t
\end{equation}
\\
The above specification is known as \emph{instantaneous} Taylor rule. However, we can suppose that the interest rate $i_t$ slowly adjusts to the target. An example of this adjustment process can be found in \citeauthor{molodtsova2009}, where:

\begin{equation}\label{eq:9}
i_t = (1-\rho) \hat{i}_t  + \rho i_{t-1} + \epsilon_t
\end{equation}
\\
Supposing that the Eq. \ref{eq:9} is applied to the data of a foreign country, we estimate:

\begin{equation}\label{eq:10}
\Delta s_{t} = \alpha + \beta_1 (\pi_t - \pi^*_t) + \beta_2 (y^{gap}_t -y^{*gap}_t) + \beta_3 ( i_{t-1} -  i^*_{t-1}) + \epsilon_t
\end{equation}
\\
that is defined as a Taylor rule \emph{with smoothing}. The presence of smoothing reflects the assumption made about the adjustment mechanism to the interest rate target. Using Taylor rule fundamentals as predictors, \citeauthor{molodtsova2009} found that the out-of-sample exchange rates forecasts are significantly better than the random walk model for several countries. Other studies \citep[e.g.][]{molodtsova2011, giacomini2010, rossi2012} also found evidence in favour of Taylor rule fundamentals. On the other hand, \cite{rogoff2008} found that the empirical evidence in favour of this fundamentals is not robust, assessing that the Taylor rule framework is a good description of monetary policies only for the past three decades. Nowadays, after the financial crises, monetary policies have been changed. It is interesting to highlight that \citeauthor{rogoff2008} analysed quarterly data instead of monthly data, as well as \citeauthor{molodtsova2009}, \citeauthor{molodtsova2011} and \citeauthor{rossi2012}.

\section{Statistical methodology}


Traditional literature of exchange rate forecasting implemented standard statistical models that incorporate economic predictors \citep{meese1983, meese1988}. These statistical models are mainly based on single equations within a linear regression framework, where the estimation of the relationships showed in Section 2 are done by \emph{ordinary least squares} (OLS). This approach has been followed, for example, by \cite{cheung2005, bacchetta2009, ferraro2015}. Alternatively, some authors proposed to include some lags, fitting a \emph{distributed lag model} \citep[e.g.][]{wright2008,molodtsova2009, molodtsova2011}. Moreover, in the class of single-equation models, another widely used alternative is the \emph{error correction model} (ECM), which assumes a long-run relationship between exchange rate levels and predictor levels.

The co-integration vector parameter can be either calibrated \citep[e.g.][]{mark1995, chinn1995, abhyankar2005, berkowitz2001long, kilian1999} or estimated \citep[e.g.][]{alquist2008, chinn2012, cheung2005, cheung2019}. Positive evidence favouring the ECM model within a long time horizon has been found by \citeauthor{mark1995}, whereas most of the other authors find no predictive ability. More interestingly $-$ using exactly the same ECM specification of \citeauthor{mark1995} $-$ \citeauthor{kilian1999}, \cite{groen1999} and \cite{groen2002} find no predictive ability for monetary models. In other words, single-equation models without a co-integrating relation provide better out-of-sample forecasts for exchange rates. For this reason, in the following, we focused on single-equation models involving mixed-frequencies \citep{ghysels2004, ghysels2007}. 

The main problem related to the usual single-equation approaches is that mixed frequencies are not allowed, even if we know that the sampling frequency affects forecasting accuracy \citep{silva2018}. As discussed by \cite{rossi2013} about estimating the equations with quarterly data, the majority of previous papers did not find favourable evidence for the classical economic predictors, differently to the studies that used monthly data (e.g. \citeauthor{molodtsova2009}). As we stated initially, the Meese and Rogoff puzzle could be potentially explained by temporal aggregation biases that rise in aggregating monthly data to a quarterly frequency. Because of temporal aggregation, much information is lost \citep{marcellino1999}. To avoid the consequences of the bias, a statistical model that allows incorporating all the monthly information available in the data should be preferred. In exchange rate forecasting, temporal aggregation is a choice related to the dependent variable, which is available weekly or daily or at higher frequencies. Instead, macroeconomic fundamentals like the interest rates, the price levels or the monetary aggregates are all monthly available at the \emph{highest} frequency\footnote{In some cases there are good high-frequency proxies for low-frequency variables. An example is quarterly GDP that could be replaced, under certain circumstances, by the monthly Industrial Production \citep[see][]{mattera2020}.}. Thus, differently from the dependent variable, temporal aggregation for the predictors is not a researcher choice.

In the class of single-equation models, \emph{mixed data sampling} \citep[MIDAS:][]{foroni2015} regression seems to be very promising in facing this kind of problems. MIDAS shares some features with distributed lag models, and from several point of views they are very similar. The basic single equation with high-frequency regressor and low frequency dependent variable is:

\begin{equation}\label{eq:11}
y_t = \beta_0 + \beta_1 B(L^{1/m};\theta)x^{(m)}_t+\epsilon^{(m)}_t
\end{equation}
\\
where $(L^{1/m};\theta) = \sum_{k=0}^{K}B(k;\theta)L^{k/m}$ and $L^{1/m}$ is a lag operator such that $L^{1/m}x^{(m)}_t = x^{(m)}_{\frac{t-1}{m}}$. 
(for $t=1,...,T$). We suppose that $y_t$ is observed at low frequency (e.g. quarterly) and $x^{(m)}_t$ is observed $m$ times in the same period (suppose $x^{(m)}_t$ is monthly or $m=4$). Therefore, it is clear that we are projecting $y_t$ onto a history of lagged observation of the high-frequency variable $x^{(m)}_{t-k}$. The parameterisation of the lagged coefficients of $B(k;\theta)$ in a parsimonious way is one of the MIDAS key features that avoid parameter proliferation. Diverse are the choices for $B(k;\theta)$ but the most common are the exponential Almon lag and Beta function \citep{ghysels2007}.

\citeauthor{foroni2015} introduced also the so-called \emph{unrestricted} MIDAS (U-MIDAS) which has very appealing features. As the authors showed in their study, when the difference in sampling frequencies between the dependent variable and the regressors is not so large (as often happen with macroeconomic applications), it might not be necessary to employ distributed lag functions $B(k;\theta)$. The essential operation made in estimating equations within the MIDAS framework is the so-called \emph{temporal alignment}. The frequency alignment is used to transform an high-frequency vector $\mathbf{x}$ with $mT$ elements into a low-frequency matrix $\mathbf{X}$ with $T$ rows and $m$ columns known as \emph{stacked vectors}:\\


\begin{equation}
\mathbf{x} =
\begin{sbmatrix}\\
x_{1} \\
\vdots \\
x_{(mT)}
\end{sbmatrix}
\rightarrow
\begin{sbmatrix}\\
x_{m} & x_{m-1} & \dots & x_1\\
\vdots & \vdots & \vdots & \vdots \\
x_{(mT)} & x_{(mT-1)} & \dots & x_{(mT-(m-1))}
\end{sbmatrix}
= \mathbf{X}
\end{equation} 
\smallskip

The MIDAS mapping follows a simple time-ordering aggregation scheme. Suppose that $y_t$ is observed quarterly and the aim is to explain its relationship with the monthly-observed variable $x_t$. Stated that each quarter has three months, a value of $m=3$ has to be used. Let consider that only the monthly data in the current quarter have explanatory power (i.e. we are estimating a single equation without lags). Assuming that the exchange rates are quarterly observed, it is possible to transform a high-frequency predictor $\mathbf{x}$ in a low-frequency matrix $\mathbf{X}$ with $m=3$ stacked vectors:\\

\begin{equation}
\mathbf{x} =
\begin{sbmatrix}\\
x_{1} \\
x_{2} \\
\vdots \\
x_{(3T)}
\end{sbmatrix}
\rightarrow
\begin{sbmatrix}\\
x_{3} & x_{2} & x_{1}\\
x_{6} & x_{5} & x_{4}\\
\vdots & \vdots & \vdots \\
x_{(3T)} & x_{(3T-1)} & x_{(3T-2)}
\end{sbmatrix}
= \mathbf{X}
\end{equation} 
\smallskip

The previous formalisation indicates that for the quarter $t$ we want to model $y_t$ as a linear combination of the monthly predictors observed within each quarter $t$. Alternatively, we can write:

\begin{equation}\label{eq:12}
y_t = \alpha + \beta_1 x_{3t} + \beta_2 x_{3t-1} + \beta_3 x_{3t-2} + \epsilon_t
\end{equation}
\\
The frequency alignment procedure turns a MIDAS regression into a classical time series regression where all the variables are observed at the same frequency. Moreover, this operation makes the single equation model to be estimated by OLS. An important advantage is that with MIDAS techniques we use all the available information for predicting the subsequent quarter. Moreover, MIDAS regression is very promising in explaining the role of temporal aggregation bias in exchange rate predictions. Therefore, we propose the mixed-frequency extensions of the models presented in the Section 2. For example, we can estimate the mixed frequency UIRP model as in the following:

\begin{equation}
\Delta s_{t} = \alpha + \beta_1 (i_{3t}- i^*_{3t}) +  \beta_2 (i_{3t-1}- i^*_{3t-1}) +  \beta_3 (i_{3t-2}- i^*_{3t-2}) + \epsilon_t
\end{equation}
\\
where $(i_{3t}- i^*_{3t})$, $(i_{3t-1}- i^*_{3t-1})$ and $(i_{3t-2}- i^*_{3t-2})$ are the inter-quarterly interest rates differences. In a similar fashion, we can extend the PPP model using mixed frequencies:

\begin{equation}
\Delta s_{t} = \alpha + \beta_1 (p_{3t}- p^*_{3t}) +  \beta_2 (p_{3t-1}- p^*_{3t-1}) +  \beta_3 (p_{3t-2}-p^*_{3t-2}) + \epsilon_t
\end{equation}
\\
where $(p_{3t}- p^*_{3t})$, $(p_{3t-1}- p^*_{3t-1})$ and $(p_{3t-2}- p^*_{3t-2})$ are the inter-quarterly price levels' differences. The same specification for the monetary models (\ref{eq:3}) and (\ref{eq:4}) produces:

\begin{align}
\Delta s_{t} &= \alpha + \beta_1 (i_{3t}- i^*_{3t}) +  \beta_2 (i_{3t-1}- i^*_{3t-1}) +  \beta_3 (i_{3t-2}- i^*_{3t-2}) + \nonumber\\
& + \gamma_1 (y_{3t} - y^*_{3t}) + \gamma_2 (y_{3t-1} - y^*_{3t-1}) + \gamma_3 (y_{3t-2} - y^*_{3t-2}) + \nonumber \\
&  + \delta_1 (m_{3t}- m^*_{3t}) +  \delta_2 (m_{3t-1}- m^*_{3t-1}) +  \delta_3 (m_{3t-2}- m^*_{3t-2}) + \epsilon_t \\
\nonumber \\
\Delta s_{t} &= \alpha + \beta_1 (i_{3t}- i^*_{3t}) +  \beta_2 (i_{3t-1}- i^*_{3t-1}) +  \beta_3 (i_{3t-2}- i^*_{3t-2}) + \nonumber\\
&+ \gamma_1 (y_{3t} - y^*_{3t}) + \gamma_2 (y_{3t-1} - y^*_{3t-1}) + \gamma_3 (y_{3t-2} - y^*_{3t-2}) + \nonumber\\
& + \delta_1 (m_{3t}- m^*_{3t}) +  \delta_2 (m_{3t-1}- m^*_{3t-1}) +  \delta_3 (m_{3t-2}- m^*_{3t-2}) + \nonumber \\
&+ \zeta_1 (p_{3t} - p^*_{3t}) + \zeta_2 (p_{3t-1} - p^*_{3t-1}) + \zeta_3 (p_{3t-2} - ^*_{3t-2}) + \epsilon_t
\end{align}
\\
In the case of \emph{instantaneous Taylor rule}, it easily follows that:

\begin{align}
\Delta s_{t} = \alpha + \beta_1 (\pi_{3t}-\pi^*_{3t}) + \beta_2 (\pi_{3t-1}-\pi^*_{3t-1}) + \beta_3  (\pi_{3t-2}-\pi^*_{3t-2}) + \gamma (y^{gap}_{t} - y^{*gap}_{t})  + \epsilon_t
\end{align}
\\
An inter-quarterly adjustment mechanism for the interest rate can be supposed to be:

\begin{equation}
i_{3t} = (1- \rho_1 - \rho_2 - \rho_3) \hat{i}_{t} + \rho_1 i_{3t} + \rho_2 i_{3t-1} + \rho_3 i_{3t-2} + \epsilon_t
\end{equation}
\\
where $\hat{i}_{t}$ is the quarterly target level of the interest rate, $i_{3t}$ is the interest rate in the end of the quarter, $i_{3t-2}$ the second month of the quarter and $i_{3t-1}$ the first one. To test the validity of the \emph{Taylor rule with inter-quarterly smoothing} mechanism of the interest rate, we can finally estimate the following relation:

\begin{align}
\Delta s_{t} = \alpha + \beta_1 (\pi_{3t}-\pi^*_{3t}) + \beta_2 (\pi_{3t-1}-\pi^*_{3t-1}) + \beta_3  (\pi_{3t-2}-\pi^*_{3t-2}) + \nonumber\\
+ \gamma (y^{gap}_{3t} - y^{*gap}_{t}) + \delta_1 (i_{3t}-i^*_{3t}) + \delta_2 (i_{3t-1}-i^*_{3t-1}) + \delta_3 (i_{3t-2}-i^*_{3t-2}) + \epsilon_t
\end{align} 

\section{An application to CAD/USD exchange rate}

\begin{figure}[!htbp]
	\centering
	\includegraphics[width=\linewidth]{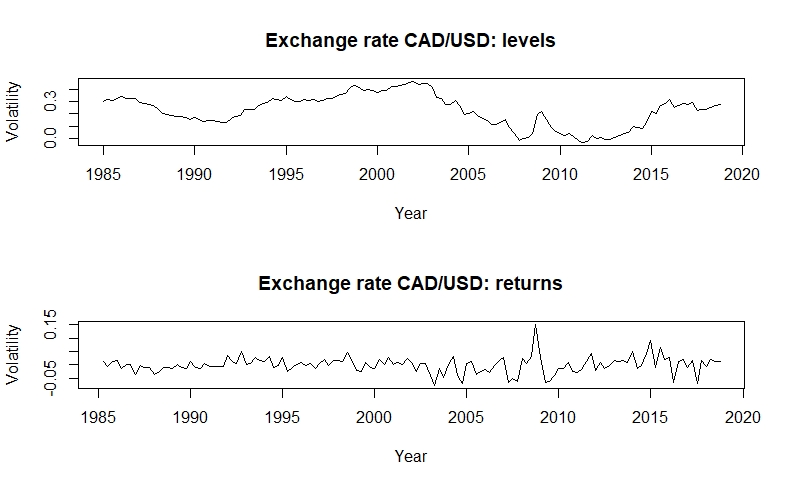}
	\caption{Exchange rate CAD/USD: levels vs returns}
	\label{fig:f1}
\end{figure}

To show the forecasting ability of the proposed approach, we considered the quarterly data of the Canadian Dollar (CAD) / U.S. Dollar (USD) exchange rate. We collected data from FRED database\footnote{\url{https://fred.stlouisfed.org/tags/series}} and computed the logarithm of the nominal monthly CAD/USD exchange rate from 01/01/1985 to 01/01/2019. More recent data about 2020 were not considered because of the COVID-19 pandemic. We then aggregated the monthly data into quarterly data and calculated the returns of exchange rates (Fig. \ref{fig:f1}).

\begin{figure}[!htbp]
	\centering
	\includegraphics[width=\linewidth]{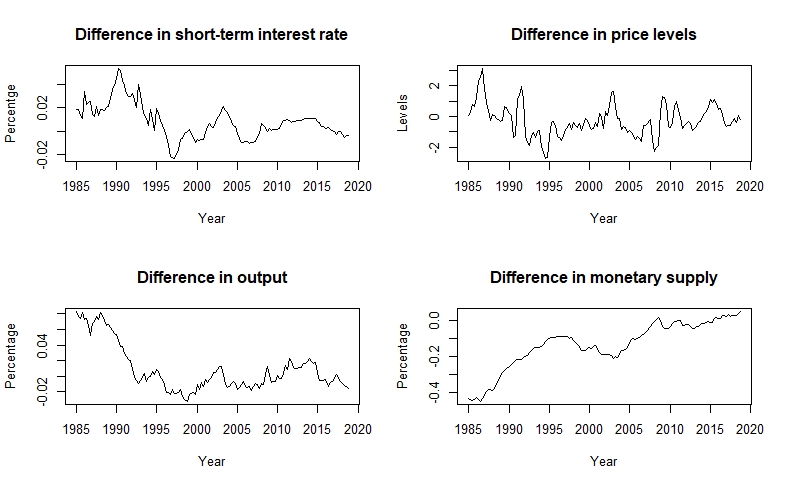}
	\caption{Relevant predictors for classical models}
	\label{fig:f2}
\end{figure}

To empirically test the performances of the UIRP-based model, we used data related to the short-term interest rate collected by the OECD database\footnote{\url{https://data.oecd.org/}} as in \cite{molodtsova2009}. The price levels, necessary to make forecasts with PPP-based predictors, are captured by the monthly logarithmic Consumer Price Index (CPI). For monetary models, we downloaded data of money supply index from the OECD database according to M3 monetary stock definition and compute the logarithm of this variable. As output variable, we considered the quarterly GDP measured in logarithmic levels. All the variables (Fig. \ref{fig:f2}) are expressed as the differences between the domestic (Canada) and foreign (U.S.) countries' values.

\begin{figure}[!htbp]
	\centering
	\includegraphics[width=\linewidth]{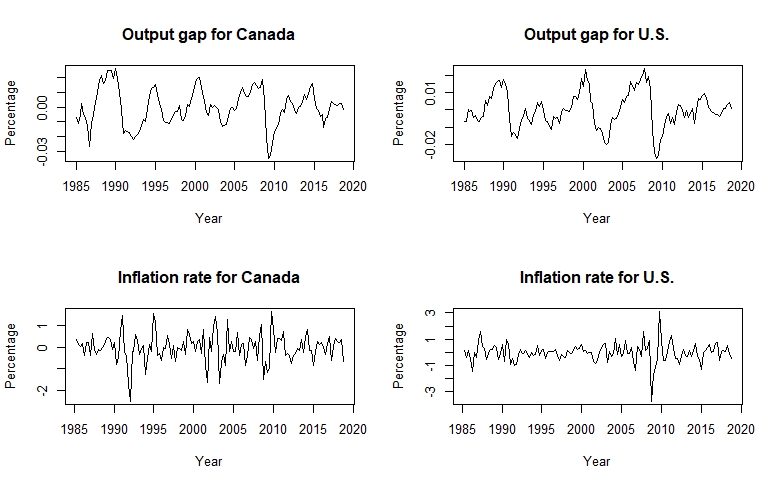}
	\caption{Other predictors}
	\label{fig:f3}
\end{figure}

Moreover, we need the output gap for the implementation of Taylor rule-based models. Following the literature, we computed the output gap as the GDP deviation from its long-run trend, obtained by applying the \cite{hodrick1997} filter. The country's inflation rate are computed as the first difference of the price levels logarithm (the predictors are shown in the Fig. \ref{fig:f3}).

We evaluated the performances of the standard models as well as of the proposed mixed frequency approach. Following \cite{ramzan2012} and \cite{chung2017}, we compared model performances with an out-of-sample analysis. We considered both a recursive approach, where the sample has an increasing size, and a rolling-window approach with a fixed sample size. To evaluate the forecasting accuracy, we used the Mean Square Forecast Error (MSFE) defined as:

\begin{equation}
\text{MSFE} = \sum_{n} (\widehat{\Delta s_t} - \Delta s_t)^2/n
\end{equation}
%
%
%

\noindent where, in general, the quantity $\widehat{\Delta s_t} - \Delta s_t$ represents the forecast error. The model with the lowest value of the associated loss function is the best one. 

Since assessing the accuracy's improvement is not enough, it is necessary to test that the forecasts obtained with alternative models are statistically different. There are several possible approaches for this purpose. The primary predictive accuracy test in the forecasting literature is the Diebold-Mariano test \citep{diebold2002}. Given two alternative forecasting models $i$ and $j$, and considering a generic loss function $g(\epsilon_{i,t})$, the loss difference is computed as in the following:

\begin{equation}
d_t = g(\epsilon_{i,t}) - g(\epsilon_{j,t})
\end{equation}
\\
The null hypothesis of the test is:

\begin{equation*}
H_0 : E(d_t) = 0
\end{equation*}
\\
where $d_t$ follows a $N(0,1)$ distribution. However, the main drawback of this approach $-$ as pointed out by \cite{diebold2015} $-$ is that the test compares forecasts but does not compare models. Therefore, following the exchange rate forecasting literature, we also used the \cite{clark2006} test to compare the performances of the proposed complex models. 

\section{Empirical results and discussion}

According to the unit root tests of \cite{said1984} (ADF) and \cite{kwiatkowski1992} (KPSS), we obtained evidence of stationarity for all the considered variables (see Table \ref{tab:t1}). For the ADF test, we considered a null hypothesis of not stationarity, while for KPSS we considered a null hypothesis of stationarity. To get consistent estimates, stationarity of all the involved variables is required. The integration order represents the number of differentiations that the time series need in order to be covariance-stationarity according to both tests.

\begin{table}[!htbp]
	\centering
	\begin{threeparttable}
		\caption{Unit root test results}
		\begin{tabular}[width=\textwidth]{l l l c}
			\hline
			\textbf{} & {ADF} & {KPSS} & {Integration order}\\
			\hline
			CAD/US exchange rate & -4.2404*** & 0.1205 & $I(0)$ \\
			Interest rate differential & -2.4043 & 0.9066*** & $I(1)$\\
			Price level differential	& -3.1993* & 0.2240 & $I(0)$\\
			Money supply differential & -2.8085 & 2.1546*** & $I(1)$ \\
			Output differentials & -2.0744 & 1.1409*** & $I(1)$ \\
			Canada output gap & -4.7458*** & 0.0337 &  $I(0)$\\
			U.S. output gap & -4.1525*** & 0.0362 & $I(0)$\\
			Canada inflation rate & -5.5314*** & 0.0324 & $I(0)$\\
			U.S. inflation rate & -5.2739*** &  0.0221 & $I(0)$\\
			\hline
		\end{tabular}
		\label{tab:t1}
		\begin{tablenotes}
			\small
			\item *** significance at 1\%, ** at 5\% and * at 10\%
		\end{tablenotes}
	\end{threeparttable}
\end{table}

To make the out-of-sample analysis, the first step is to split the sample into a training set and a testing set. We obtained the forecasts according to both a recursive and a rolling window schemes. As training set we selected the period from 1985 to 1994, while as testing set we selected the time window between 1995 and 2019. Table \ref{tab:t2} contains the list of the estimated models in our empirical analysis.

The results of the forecasting accuracy obtained by the recursive scheme and the associated predictive accuracy are in Table \ref{tab:t3}. The majority of the forecasting models (the only exception is the MM$_2$ model) provided more accurate forecasts than the benchmark in Eq. \ref{eq:0}. We considered a random walk model without drift. Forecasts are one step ahead ($h=1$). Under the DM column are reported the value associated with the \citeauthor{diebold2002} test, computed assuming MSFE loss function. Under the CW column, instead, is reported the value associated with the \citeauthor{clark2006} test.

We observed a better performance of the Taylor rule-based models. In particular, we achieved the most accurate forecasts with the instantaneous Taylor rule model (TYLR$_1$), whereas the Taylor rule model with smoothing (TYLR$_2$) performed poorer. The worst model seemed to be the sticky price version of the monetary model (MM$_2$). The Taylor rule with inter-quarterly adjustment (MF-TYLR$_2$), proposed in this paper, provided the most accurate forecasts within the class of mixed-frequency models. 

\begin{table}[!htbp]
		\centering
 \begin{threeparttable}
	\caption{List of estimated models}
	\begin{tabular}[width=\textwidth]{l l}
		\hline
		\textbf{Acronym} & \textbf{Model description}\\
		\hline
		UIRP &  Uncovered Interest Rate Parity estimated by the equation (2)\\
		\hline
		PPP & Purchasing Power Parity estimated by the equation (3)\\
		\hline
		MM$_1$ & Flexible price monetary model estimated by the equation (4)\\
		\hline
		MM$_2$ & Sticky price monetary model estimated by the equation (5)\\
		\hline
	        TYLR$_1$ & Instantaneous Taylor rule estimated by the equation (9) \\
		\hline
		TYLR$_2$ & Taylor rule with smoothing estimated by the equation (11) \\
		\hline
	        MF-UIRP &  Mixed frequency version of (2) estimated by (16) \\
		\hline
	        MF-PPP &  Mixed frequency version of (3) estimated by (17)  \\
		\hline
                MF-MM$_1$ & Mixed frequency version of (4) estimated by (18) \\
		\hline
		MF-MM$_2$  & Mixed frequency version of (5) estimated by (19) \\
		\hline
		 MF-TLYR$_1$ & Mixed frequency version of (9) estimated by (20) \\
		 \hline
	         MF-TYLR$_2$ &  Mixed frequency version of (11) estimated by (22) \\
		\hline
		\end{tabular}
		\label{tab:t2}
	 \end{threeparttable}
		\end{table}

\begin{table}[!htbp]
		\centering
	 \begin{threeparttable}
			\caption{Out-of-sample analysis: recursive approach}
			\begin{tabular}[width=\textwidth]{l l l l}
				\hline
				\textbf{} & {MSFE} & {DM} & {CW}\\
				\hline
				Random Walk & 0.001984 & - & - \\
				\hline
				UIRP & 0.001649 & 7.7800*** & 1.53e$^{-18}$*** \\
				PPP & 0.001776 & 6.6951*** & 5.87e$^{-13}$*** \\
				MM$_1$	& 0.001500 & 7.3806*** & 8.98e$^{-19}$***\\
				MM$_2$ &  0.003241 & -7.6012*** & 1.01e$^{-16}$*** \\
				TYLR$_1$ & 0.001155 & 3.2853*** &  9.71$^{-17}$***\\
				TYLR$_2$ & 0.001547 & 7.3827*** & 1.12e$^{-18}$***\\
				MF-UIRP & 0.001612 & 7.7164*** & 8.21e$^{-19}$***\\
				MF-PPP & 0.001630 &  7.5522*** & 3.28e$^{-16}$***\\
				MF-MM$_1$ & 0.001439 &  6.9542*** & 3.15e$^{-18}$*** \\
				MF-MM$_2$ & 0.001504 & 7.1243*** & 5.87e$^{-18}$***\\
				MF-TYLR$_1$ & 0.001318 & 6.0342*** & 3.39e$^{-16}$***\\
				MF-TYLR$_2$ & 0.001266 & 5.8580*** & 1.40e$^{-18}$***\\
				\hline
			\end{tabular}
			\label{tab:t3}
			\begin{tablenotes}
				\small
				\item *** significance at 1\%, ** at 5\% and * at 10\%
			\end{tablenotes}
		 \end{threeparttable}
		\end{table}

In conclusion, with a recursive approach, the mixed frequency based extensions improved the forecasting accuracy in comparison with the classical models. In particular, the MF-UIRP was the 2.3\% more accurate than UIRP, the MF-PPP was the 8.2\% more accurate than PPP, and the MF-TYLR$_2$ was the 18\% more accurate than TYLR$_2$. The highest benefit of considering a mixed-frequency model was reached, in terms of accuracy, with the mixed-frequency sticky price version of the monetary model (MF-MM$_2$), that was the 53.6\% more accurate than the classical MM$_2$ model.

In a similar fashion, we evaluated the forecasting accuracy of the proposed models according to a rolling window scheme (Table \ref{tab:t4}).

\begin{table}[!htbp]
				\centering 
	\begin{threeparttable}
			\caption{Out-of-sample analysis: rolling w. approach}
			\begin{tabular}[width=\textwidth]{l l l l}
				\hline
				\textbf{} & {MSFE} & {DM} & {CW}\\
				\hline
				Random Walk & 0.001984 & - & - \\
				\hline
			UIRP & 0.002137 & -1.2658 & 0.1498 \\
			PPP & 0.001596 & 7.1651*** & 2.73e$^{-15}$*** \\
			MM$_1$	& 0.002355 & -2.0977*** & 0.1238\\
			MM$_2$ & 0.005823 & -7.1600*** & 0.9999 \\
			TYLR$_1$ & 0.001189 & 4.2468*** &  1.25e$^{-17}$***\\
			TYLR$_2$ &  0.001550 & 5.2236*** & 8.15e$^{-11}$***\\
			MF-UIRP & 0.001923 & 0.5376 & 0.001346***\\
			MF-PPP & 0.001482 &  7.1987*** & 9.84e$^{-18}$***\\
			MF-MM$_1$ &  0.005702 & -3.2726*** & 0.08108* \\
			MF-MM$_2$ & 0.007596 & -3.5049*** & 0.1251\\
			MF-TYLR$_1$ & 0.001447 & 6.6033*** & 4.21e$^{-18}$***\\
			MF-TYLR$_2$ & 0.001844 & 2.9112*** & 1.43e$^{-04}$***\\
				\hline
			\end{tabular}
			\label{tab:t4}
			\begin{tablenotes}
				\small
				\item Note: *** significance at 1\%, ** at 5\% and * at 10\%
			\end{tablenotes}
		 \end{threeparttable}
\end{table}

The results were completely different from those reported in Table \ref{tab:t3}. First of all, the presence of the \cite{meese1983} puzzle is here more evident. For example, the classical UIRP model is not able to predict the exchange rate, as well as the monetary models. This result was confirmed both by the \citeauthor{diebold2002} and the \citeauthor{clark2006} tests. As in \cite{molodtsova2009} and \cite{molodtsova2011}, the Taylor rule-based forecasts provided good results, since both the instantaneous rule (TYLR$_1$) and the smoothing rule (TYLR$_2$) provided better forecasts than the benchmark.

As in the recursive approach, with the mixed-frequency extensions we obtained better results than the classical models for the majority of the cases. The MF-UIRP model provided better results than the benchmark in terms of MSFE (the classical UIRP specification based on temporal aggregation provided less accurate forecasts). The gain in accuracy of the MF-UIRP with respect the classical UIRP was exactly equal to 10\%. While the classical UIRP provided the same forecasts of the random walk without drift, with the MF-UIRP specification we obtained a significant over performance. In other words, the unpredictability of UIRP model was explained by temporal aggregation. Similar conclusions can be drawn for the MF-PPP model, since it provided more accurate forecasts than the classical PPP model with an accuracy gain close to 7.2\%.

The overall evidence of the presented study suggests different findings with respect to other recent studies (e.g. \cite{cheung2019}), in which there are evidence against predictability for CAD/USD exchange rate using quarterly data. An important result is that by incorporating mixed frequency we were able to improve forecasting accuracy.
		
%
%

\section{Conclusions}

According to the economic theory, several variables can be used to explain the exchange rate fluctuations. From an empirical viewpoint, a vast literature argued instead that the variables used in the most popular frameworks are not able to forecast the exchange rate better than a simple random walk model. This result is called the \emph{Meese and Rogoff puzzle}. The most shared explanation of the puzzle is that, stated the validity of the economic theory, the unpredictability should derive by the presence of sampling errors or to statistical models' misspecification.

The most common statistical approach to exchange rate forecasting is based on the classical linear regression model. Starting from the work of \cite{mark1995}, several authors tried to incorporate the long-run relationships among the predictors as additional variables, with poorer results than standard linear regression model. In this paper, we claimed that a possible explanation of the \citeauthor{meese1983} puzzle can be found in the so called \emph{temporal aggregation biases}. These biases, caused by the information lost induced by temporal aggregation, can be seen as a source of misspecification when some important (high-frequency) variables are omitted from the model. This intuition lies on the fact that the results presented in the literature are clearly affected by the frequency at which exchange rates are sampled. Even thought exchange rate data are daily available, many studies focused on monthly or quarterly frequencies \citep{rossi2013}.

The mixed-frequency regression model is a well known technique able to overcome this issue. Here we proposed for the first time, to the best of our knowledge, to use the monthly-sampled predictors to forecast the (long-run) quarterly exchange rates by means of a Mixed Data Sampling (MIDAS) regression. The main empirical finding of the paper $-$ on the basis of a case study concerning the CAD/USD exchange rate $-$ is that the mixed-frequency regression model improves the predictive ability in comparison with the classical models, that are instead affected by the temporal aggregation bias. Therefore, the contribution of this paper is two-fold. First of all, we showed the implementation of the MIDAS regression to predict quarterly exchange rates with very promising results, offering a new applicative domain for this approach. Second of all, we provided a possible explanation of the \citeauthor{meese1983} puzzle. These findings can be interesting for a varied audience, including both scholars and practitioners.


\bibliographystyle{apalike}
\bibliography{reference}

\end{document}